# AGE BASED USER INTERFACE IN MOBILE OPERATING SYSTEM

Sumit Sharma[1], Paramjit Singh[2], Rohitt Sharma[3] and Aditya Mahajan[4]

[1]Research Scholar, Department of Computer Science Engineering, Lovely Professional University, Phagwara, India

`Sumit_sharma@mailingaddress.org`

[2]Assistant Professor, Department of Computer Science Engineering, Lovely Professional University, Phagwara, India

`paramjit.12969@lpu.co.in`

[3]Research Scholar, Department of Computer Science Engineering, Lovely Professional University, Phagwara, India

`rohittsharma@mailingaddress.org`

[4]Graduate Engineer, Department of Computer Science Engineering, Lovely Professional University, Phagwara, India

`adityamahajan3@gmail.com`

## ABSTRACT

*The mobile phones are becoming now an irreplaceable utility of every household. It serves as wall clock, alarm clock, calculator, calendar, timer and many more, but have this multi-functionality overloaded the interface of the new generation of the mobile phones. The youth have adapted well to these multiple functionalities graphical user interface, but the interface has now haunting effects for usage by the two age groups i.e. elderly and kids. The interface may end up leading the new generation mobiles in the market useless or of very little use to the elderly and kids. This leads towards the need of age based user interface in the mobile operating system which will consist of interface selection home screen which further directs to age oriented interfaces.*

## 1. INTRODUCTION

In today's world, the market is pullulating with powerful communication units embedded with highly-advanced technologies. The importance of the mobile phone in human creed could be seen from the fact in the report of CISCO that **by the end of 2012, the number of mobile-connected devices will exceed the number of people on earth, and by 2016 there will be 1.4 mobile devices per capita**. There will be over 10 billion mobile-connected devices in 2016, including machine-to-machine (M2M) modules-exceeding the world's population at that time (7.3 billion) [1].

Mobile phones eliminated the main disadvantage of telephone that is immobility. But this isn't the only thing mobile phone are capable of. Mobile phones now act as a single source multiple usage devices. Speaking of smart phones, these devices are the most popular among mobile phone types. Due to their multimedia capabilities, smart phones extend the importance and utility of mobile phones not only as a reliable communication device but as a compact entertainment gadget, as well.

The leading brands in the market of smart phones are: Apple, Nokia, Android, Blackberry and Samsung. Seeing that the number of mobile phones will outnumber the population of earth there is certain need that the mobile phones must evolve with new offerings to the user to keep up in the stiff competition in the market. So, this field has got lot of opportunities to be explored. The rest of paper is organised as follows: section 2 explains user interface its types and cellular phones' user interface. Section 3 discusses the Touch Screen Approach which is dominant in the form factors among all smartphones and also discusses the two types of touch response mechanisms. Section 4 describes the research methodology used. Section 5 inspects the factors affecting the user interface acceptance. Section6 helps understand the user prospective canvass. Section 7 concludes the results obtained and proposes the approach to achieve the inferences generated in the previous questions.

## 2. USER INTERFACE

User Interface is the component of a product where human interacts with the product. A *user interface* is a linkage between a human and a device or system that allows the human to interact with (e.g., exchange information with) that device or system. [2] In the domain of computers the types of User Interface available are: Command Line Interface, Graphical User Interface. The command line interface accepts requests made through keyboard only in a non-user friendly environment and displays the respective results on the terminal screen. The command line interface requires high level of expertise and is hence constrained for usage only by highly qualified professionals. The graphical user interface is a complete opposite of the command line interface. In this type of interface the user can make requests either by keyboard or mouse in a user friendly environment and provide articulated graphical output on the terminal screen.

### 2.1. Cellular Phones' User Interface

The cellular phones use only one of them i.e. Graphical User Interface. The interface in the earlier generations included only the basic functionalities of a phone on the home screen i.e. Dialler and Contact Book; but the smartphones in today's world provide a lot of widgets on the home screen and a complex interface on a whole which simplifies the world for tech savvy users but make it inversely proportional for kids and the elderly. The graphical user interface accepts input through keypad or touch screen. The graphical user interface which accepts input through touch screen is referred to as Touch User Interface.

## 3. TOUCH SCREEN APPROACH

In mobile user interface of touch screen phones the rule that dictate the designing is that clarity outflanks density, which means that the interface should not contain a lot of elements but single element or a few elements able to provide basic functionality. But now days this rule is being implemented to the individual elements but not to the complete user interface. The applications which launch as a result of the shortcut on the screen are containing lesser elements but the home screen is crowded because of trying to make every feature accessible from the home screen. Such kind of interface is highly effective. The Touch User Interface supports two types of touch screens i.e. resistive and capacitive.

### 3.1. Resistive Touch Screen

The resistive touch screen consists of two sheets of glass where one being conductive in nature which lies under the other sheet which is resistive in nature. When a point of contact is made on the screen with the help of stylus the two screens touches each other and hence a charge is produced. Resistive touch screen generally is considered to provide lesser reactive touch response but in actual it isn't so. It is the type of response it requires from user makes the user to feel so. The user needs to press the touch screen with a stylus and not just touch the screen with stylus. For this reason the user feels it to be lesser responsive.

### 3.2. Capacitive Touch Screen

In the case of capacitive touch screen an electric charge sheet is placed on the glass. When a user touches the screen with his/her finger a static charge is produced in human body which further produces electric current which informs the operating system that an interaction is made by human and hence produces required action. As in this case the user only needs to touch the screen and not press it. So, the effort required on the user side is minimalist which makes him feel the touch screen to be more responsive.

## 4. RESEARCH METHODOLOGY

A survey is done to collect information regarding the experience of the user with the graphical user interface of smartphones. It was done using a questionnaire which was created as a web survey [3]. The questionnaire link was sent to the industry professionals, academic professionals, students and others who are using mobile phones from a long period of time. The questionnaire is reviewed by industry professionals working in the profile of quality assurance and the author himself for gathering data effectively. The analysis is done based on the responses received from industry professionals, academic professionals, students and others. While making this questionnaire, various key attributes were taken into account such as what is the magnitude user experience with cellular phones, what is the age group of the effected, and what are the changes required by the users. The industry professional belonged to various IT industries listed as Headstrong Services Private Ltd., Tavant Technologies, morpho e-documents global, AON Hewitt, Infosys, J.H.C, Avant grade digital, Tata Consultancy Services, HCL COMNET, Trident India, Xchanging technologies, Syscom corporation Limited, Syntel Limited, C.S.C, Accenture, Campus EAI Consortium, Convergys, CVG; and the students participated in the web survey belong to International Institute of Information Technology(Bangalore) and Lovely Professional University. The academic professionals belong to Lovely Professional University.

## 5. TOUCH USER INTERFACE ACCEPTANCE

Acceptance of touch user interface could be determined on two bases i.e. internationally and across different ages. For making an interface acceptable throughout the world, the major factors to be considered are Text used and Formats [4]. The text used in the touch screen interface should not contain any slang words, ambiguous words or characters which are not included in the International standards. For example, if the interface contains word pomme then for French people it would mean apple but for Germans it would mean chips. So, text used in the interface should be carefully accepted. Similarly, the formats belonging to one culture could end up giving completely different.

Acceptance of user interface is also different across different ages. The dense touch user interface is acceptable to people falling in the age slab of 10-50 years. But dense touch user interface is not at all acceptable to the age group of 50+ and 10-. This is due to the fact that generally elderly people suffer from some visual impairments or a new bee for a smartphone which makes the dense touch user interface not acceptable. Similarly for kids the dense touch user interface may be too confusing or not to be related to. So, these factors and complex linkage between different applications and even within an application should be taken care of supporting the acceptance of Touch User Interface.

## 6. USER PROSPECTIVE CANVASS

The web survey conducted on three sections of society i.e. Industry Professional, Academic Professional and Students lead towards drawing some inferences. The drawn inferences will be discussed in the next section. This section discusses the questions asked to the users along with the motive behind it. The following image shows that how experienced are the respondents of the survey.

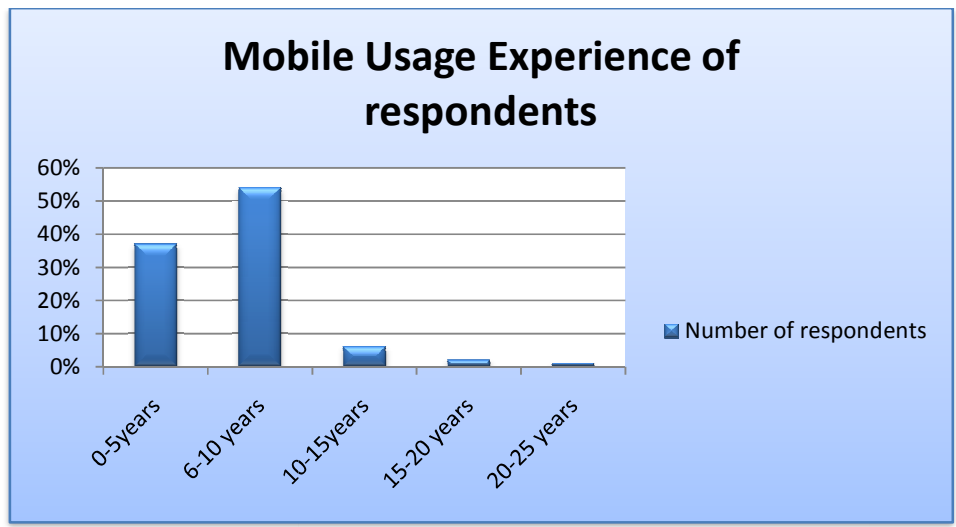

Figure 1. Respondent experience with cellular phones

This question was asked to check whether the respondents are experienced enough to be a participant in mobile phones survey. Seeing that majority of users had been using cellular phones for more than 5 years this makes their responses credible for drawing some inferences.
Now the root need of creating different interfaces was examined in the next question

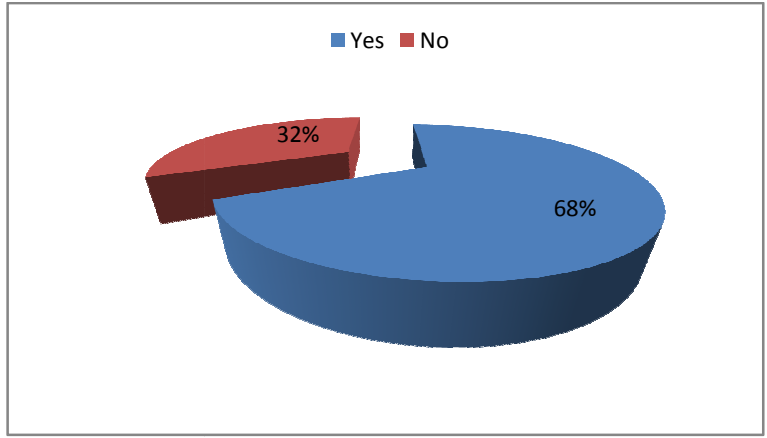

Figure 2.  Have to lend phone to others?

As from the pie chart above we can see that 68% of the respondents agreed that they had to lend their phones in the social circle or family. This makes purpose of our proposal justified i.e. as most of the users had to lend their phones, and the borrower may include people of different ages. The next important question asked was that whether their parents felt comfortable with the interface of their smartphones.

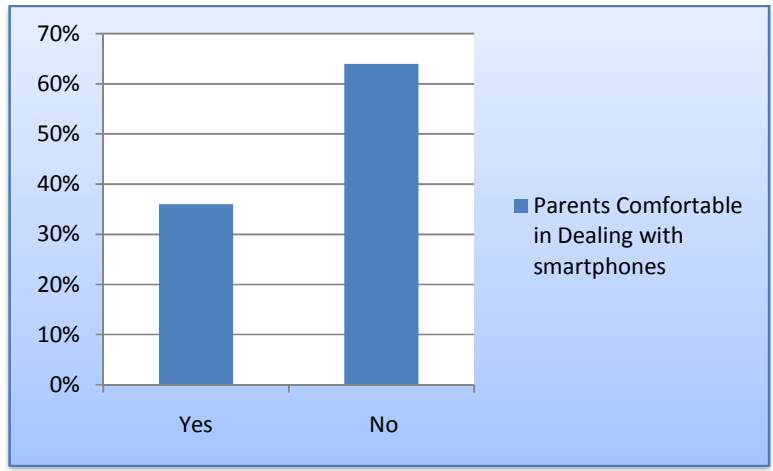

Figure 3.  Interaction Experience of Elderly

This result helps ascertain that the elderly people are finding it difficult to use the smartphones available in the market.

The subsequent important questions aimed at finding out whether the kids have created some problems to the owner by deleting some files or did some unexpected things due to complex user interface. This question is supporting the need of age based interface along with the previous question.

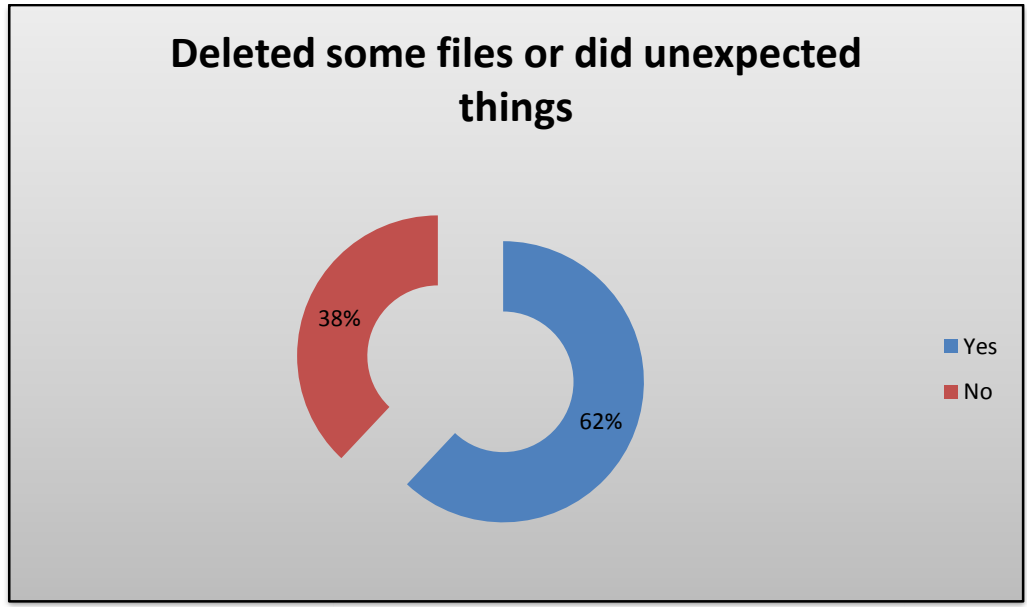

Figure 4. Interaction Experience of Elderly

The results showed that in majority of cases i.e. 62%; the owners when lends their phone to kids they had either deleted some files or did unexpected things which had created problems for the owner.

## 7. CONCLUSION

The survey lead towards results that are totally supporting the proposed issue i.e. need of age based user interface. The section of society which uses smartphones belongs majorly to industries and academic institutions. This means that our survey is covering the right subset of the population. The results indicated that elderly found it difficult to use the touch screen phones/smartphones and nearly 60% of elderly belonged to age group 40-50 years and 40% belonged to 51-60%. This leads us towards concluding that age interface for the elders should contain minimalist functionality and the icons and text should be bold as the growing age may be accompanied with low vision. The number of blind persons in the US is projected to increase by 70% to 1.6 million by 2020, with a similar rise projected for low vision. CONCLUSIONS: Blindness or low vision affects approximately 1 in 28 Americans older than 40 years.[5] The other results concluded are that kids most of the times end up doing some unexpected things when they borrow a phone. So, this leads us towards proposing a separate user interface for the kids which when entered could not be exited without some secret key. The user interface for kids should contain only games, educational widgets and music. The last result inferred is that the users are happy with their smartphones having more than one interface. This leads us towards proposing that each single device should contain all the three interfaces i.e. existing interface of smartphones, interface modelled for elders and interface created to abstract complexities of the interface from kids and providing just few applications. The following flowchart explains what we are proposing for achieving above mentioned goals.

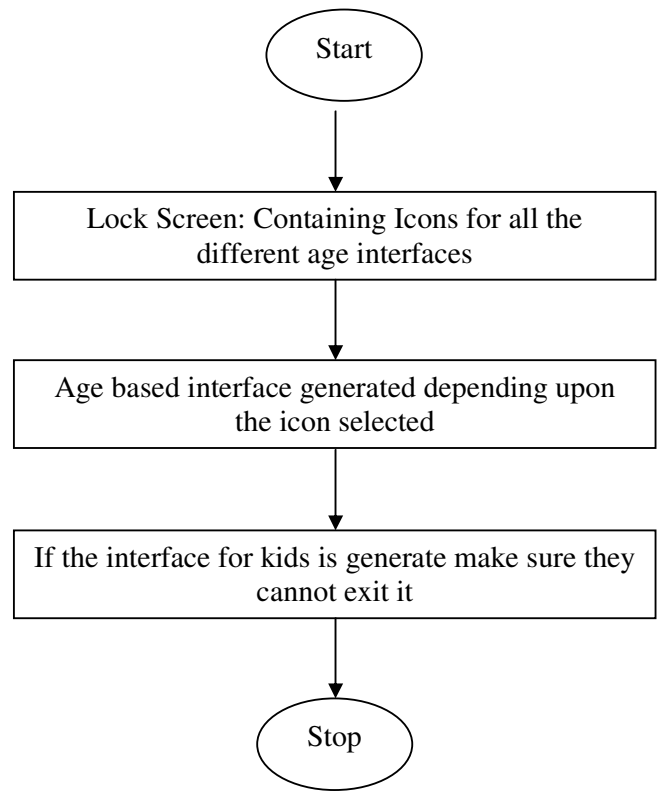

## AUTHORS

Sumit Sharma received his B.Tech degree in Computer Science from Lovely Professional University, India in 2011 and pursuing his M.Tech degree from Lovely Professional University, India. His current research interest includes Mobile Operating Systems, Software Metric Performance Analysis.

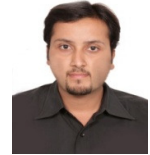

Paramjit Singh is Post Graduate in Computer Science and is currently working as Assistant Professor at Lovely Professional University. He has a teaching experie nce of around 5 years. He has 3 publications in his name including journal and conference.

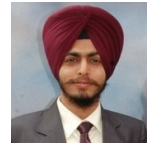

Rohitt Sharma received his B.Tech degree in Computer Science from Lovely Professional University, India in 2011 and pursuing his M.Tech degree from Lovely Professional University, India. His current research interest includes Software Metric Performance Analysis, Mobile Operating Systems.

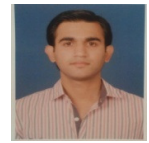

Aditya Mahajan received diploma in Cyber Crime Prosecution & Defense from Asian School of Cyber Laws, Pune in 2010 and is pursuing B.Tech from Lovely Professional University. He is Community Manager & Tech Lead of Google Technology User Group for Jammu region (March 2011 - present).

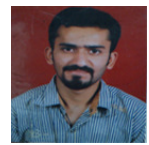